\documentclass[prl,twocolumn,showpacs,superscriptaddress]{revtex4}
\usepackage{graphicx}
\usepackage{amsmath}
\usepackage{bm}
\usepackage[T1]{fontenc}
\date{\today}
\begin{document}
\title{Probing hole-induced ferromagnetic exchange in magnetic semiconductors
by inelastic neutron scattering} 
\author{H. K\k{e}pa}
\email{Henryk.Kepa@fuw.edu.pl}
\affiliation{Institute of Experimental Physics, Warsaw University, 
Ho\.za 69, 00-681 Warszawa, Poland}
\affiliation{Physics Department, Oregon State University, Corvallis, 
OR 97331, USA}
\author{Le Van Khoi}
\affiliation{Institute of Physics, Polish Academy of
Sciences, al.~Lotnik\'ow 32/46, 02-668 Warszawa, Poland}
\author{C. M. Brown}
\affiliation{ NIST Center for Neutron Research, 100 Bureau Dr.,
Gaithersburg, MD 20899-8562, USA}
\affiliation{Department of Materials and Nuclear Engineering, University 
of Maryland, College Park, MD 20742-7531, USA}
\author{M. Sawicki}
\affiliation{Institute of Physics, Polish Academy of
Sciences, al.~Lotnik\'ow 32/46, 02-668 Warszawa, Poland}
\author{J.K. Furdyna}
\affiliation{Department of Physics, University of Notre Dame, Notre Dame, 
IN 46556, USA}
\author{T.M. Giebultowicz}
\affiliation{Physics Department, Oregon State University, Corvallis, 
OR 97331, USA}
\author{T. Dietl\footnote{Address in 2003: 
Institute of Experimental and 
Applied Physics, Regensburg University; supported by Alexander von Humboldt 
Foundation}}
\affiliation{Institute of Physics, Polish Academy of
Sciences, al.~Lotnik\'ow 32/46, 02-668 Warszawa, Poland}

\begin{abstract}
The effect of hole doping on the exchange coupling of the nearest neighbor 
(NN) Mn pairs in Zn$_{1-x}$Mn$_x$Te is probed by inelastic neutron scattering.
The difference in the NN exchange energy $\Delta J_1$ in the presence and in 
the absence of the holes is determined. The obtained value of $\Delta J_1$
is in good agreement with the predictions of the Zener/RKKY model, even on
the insulator side of the metal-insulator transition. 
\end{abstract}
\maketitle

Recent comprehensive research on the nature of carrier-controlled 
ferromagnetism in III-V and II-VI Mn-based diluted magnetic semiconductors
(DMS) has clearly shown that these systems combine intricate properties of
charge-transfer insulators and strongly correlated disordered metals with
the physics of impurity and band states in heavily doped semiconductors
\cite{Mats02Diet02}. Because of this complexity, there are diverging opinions
about the dominant microscopic mechanisms accounting for the ferromagnetism
in these materials. For instance, according to \emph{ab initio} computations,
the holes associated with the presence of Mn in III-V compounds reside in
the Mn d-band, which implies that the relevant spin-spin coupling  mechanism
is double exchange \cite{Sato02}. Other workers, guided by photoemission
results \cite{Okab98}, assume that the Mn d-band is deep in the valence band.
This means that the holes which mediate the ferromagnetic coupling originate
from effective mass acceptors produced by Mn in III-V compounds and by
extrinsic dopants, such as N, in II-VI DMS. However, even if this is the
case, a number of possible scenarios can \emph{a priori} be envisaged.
In particular, if the holes stay localized by the parent acceptors, the
ferromagnetic transition can be viewed as the percolation threshold of bound
magnetic polarons \cite{Durs02Kami02}. Alternatively, the hole-mediated
ferromagnetic coupling can be assigned to virtual transitions 
of the holes from the acceptor levels to the valence band \cite{Litv01Inou00}.
If, however, the relevant hole states are extended or their localization
length is much greater than an average distance between the acceptors,
exchange interactions will be effectively mediated by the itinerant carriers,
so that the Zener or Ruderman-Kittel-Kasuya-Yosida (RKKY) mechanisms will
operate \cite{Diet97,Diet00,Jung99Koni01}. In this situation, 
attempts to simulate the properties of these systems numerically have been
undertaken \cite{Yang03Timm02}, although it is still a formidable task to
take into account the effects of both disorder and carrier-carrier correlation
near the Mott-Anderson transition even in non-magnetic
semiconductors \cite{Beli94}.

On the experimental side, there is no evidence so-far for the d band transport
and for the associated colossal magnetoresistance in III-V and II-VI
ferromagnetic semiconductors. This seems to suggest that the double exchange
is not the dominant spin-spin coupling mechanism in these systems. On the other
hand, in both III-V \cite{Mats98} and II-VI \cite{Ferr01} Mn-based DMSs, the
ferromagnetism occurs on both sides of the metal-insulator transition (MIT) 
which, similarly to other extrinsic semiconductors, is of the Anderson-Mott
type. Interestingly, the corresponding Curie temperatures show no critical
behavior across the localization boundary \cite{Mats98,Ferr01}. It is, 
therefore, hard to tell based on macroscopic magnetization measurements
whether the holes bound by individual acceptors or, alternatively, the holes
residing in weakly localized or extended states mediate ferromagnetism in these
compounds. In order to address this question experimentally, we have examined
the magnitude of the exchange energy between the nearest neighbor (NN) Mn pairs
by inelastic neutron scattering in Zn$_{1-x}$Mn$_x$Te. By comparing results
for undoped and doped crystals we determine rather directly the contribution
to the exchange brought about by the presence of the holes. According to our
electrical measurements, the sample studied remains on the insulator side of 
the metal-to-insulator transition. Nevertheless, the hole-induced contribution
to the pair energy is by a factor of four smaller than that calculated under
the assumption that the holes reside on individual acceptors. By contrast, 
if the hole states are assumed to be metallic-like at length scales of the NN
distance, the calculated value is smaller than the experimental one by a factor
of 1.5, a discrepancy well within the combined uncertainties in input
parameters to theory and to experiment. Furthermore, no visible broadening of
the scattering line is observed in the presence of doping, again indicating
that the hole states exhibit no substantial fluctuations expected for the
strongly localized regime. We conclude that the ferromagnetic exchange is
mediated by itinerant holes, even on the insulator side of the MIT.

Because of the small magnetic cross section, inelastic scattering measurements
cannot be completed on films grown by molecular beam epitaxy commonly employed
in order to overcome the thermal equilibrium solubility limits of 
relevant dopants in DMS \cite{Mats98,Ferr01}. However, it has been shown
previously by some of us that large single crystals of Zn$_{1-x}$Mn$_x$Te:P
can be prepared \cite{Khoi02}, in which  Mn and hole concentrations are
sufficiently high to observe ferromagnetic ordering. Our  Zn$_{1-x}$Mn$_x$Te
crystals are grown by the high pressure Bridgman method from
(ZnTe)$_{1-x}$(MnTe)$_x$ solution in an evacuated ($10^{-6}$ Torr) quartz
ampoule, with phosphorus corresponding to the doping level of
$5 \times 10^{19}$ cm$^{-3}$ added in the form of Zn$_3$P$_2$. The crystals 
obtained in this way are 10--20~mm in diameter and about 60~mm long.
The x-ray energy dispersive fluorescence analysis demonstrates a uniform
distribution of Mn along the central part of the ingots. X-ray diffraction 
studies do not reveal any second phase inclusions, and shows that the lattice
constant increases linearly with the Mn content in accord with Vegard's law.  

According to Hall effect measurements carried out at room temperature, the
hole concentration in the as-grown material is about $1.3 \times 10^{19}$ in
the case of ZnTe, but decreases strongly with the Mn content, dropping down to 
$2 \times 10^{16}$ for Zn$_{0.95}$Mn$_{0.05}$Te. This suggests that the
incorporation of Mn results in the creation of rather efficient compensating
centers. A microscopic mechanism accounting for this effect is unknown at
present but since the tetrahedral radius of Mn$^{2+}$ (1.33\AA) is larger than
the tetrahedral radius of Zn$^{2+}$ (1.225 \AA) \cite{FurdKoss86}, 
interstitial Mn atoms acting as donors might be involved. It has been found,
however, that appropriate heat treatment can significantly increase the hole
concentration. In particular, annealing of Zn$_{0.95}$Mn$_{0.05}$Te under 
nitrogen pressure of 4~MPa at 800 $^{\circ}$C for a week has resulted in a
hole concentration as large as $5 \times 10^{18}$ cm$^{-3}$. Nevertheless,
according to low temperature resistance measurements such samples remain on
the insulating side of the MIT, an observation that is consistent with
previous data for Zn$_{1-x}$Mn$_x$Te:N \cite{Ferr01} and with the small value
of the acceptor Bohr radius in ZnTe, $a_B = 13$ \AA{} \cite{Ferr01}.
On the other hand, according to previous SQUID measurements, the annealed
samples exhibit the presence of ferromagnetic correlation, which leads to the
Curie-Weiss ferromagnetic temperature of about 2~K \cite{Khoi02}. 

Our inelastic neutron measurements of Mn pair spectra were carried out on a
Fermi Chopper TOF Spectrometer \cite{fcstof}
at the Center for Neutron Research at NIST.
The incident neutron energy was 3.55 meV. Two  single crystal specimens were
used: a 3 cm$^3$ doped Zn$_{0.95}$Mn$_{0.05}$Te:P and a 6 cm$^3$ undoped
Zn$_{0.98}$Mn$_{0.02}$Te as a reference sample. 
Inelastic scattering spectra were measured at 10 and 40~K. The doped sample
was investigated twice: in the as-grown, high resistivity state, and after
being annealed, as described above. 

\begin{figure}
\includegraphics[width=3in]{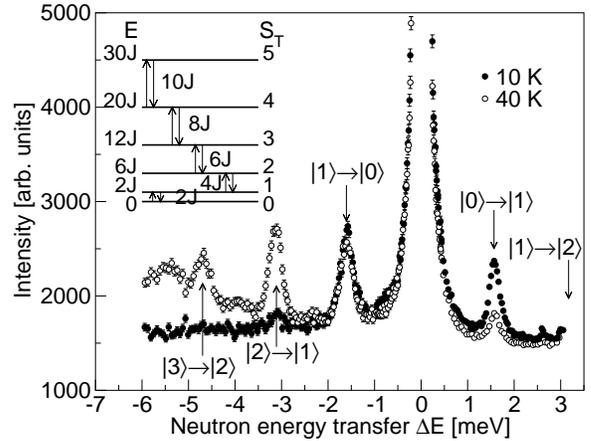}
\caption{\label{fcs1} Neutron inelastic spectra measured on the annealed
$p$-type Zn$_{0.95}$Mn$_{0.05}$Te sample at 10 and 40~K. At low temperature
(dark data points) only transitions between the ground state an the first
excited level are seen. At 40~K additional peaks corresponding to the
$|2\rangle \rightarrow |1\rangle$ and $|3\rangle \rightarrow |2\rangle$
transitions show up due to increased population of the excited states. At the
same time, the decreased  ground state population leads to the lower
$|0\rangle \rightarrow |1\rangle$ line intensity.}
\end{figure}

For a random distribution of Mn ions over the fcc cation sublattice at
$x = 0.05$, 54\% of ions are singlets, 24\% are members of 
NN pairs studied here, 4\% belong to triads, and the remainder to larger
clusters. The net value of the Mn-Mn exchange constant $J_i$ for an
$i^{\rm th}$ neighbor  shell is determined by two contributing effects --
a hole-induced ferromagnetic interaction $J_i^h$, and an intrinsic component
$J_i^{\rm int}$. The latter is known to arise from a strong but short-range
antiferromagnetic superexchange \cite{Shap02}, which is essentially insensitive
to the presence of carriers.  Thus the difference in the value of $J_1$
obtained from measurements with and without holes yields the magnitude of the
carrier-induced component $\Delta J_1^h$ that can be compared with the
predictions of theoretical models
\cite{Durs02Kami02,Diet97,Diet00,Jung99Koni01}.

An antiferromagnetically coupled ($J<0$) isolated pair of Mn$^{2+}$ spins
($S_i=S_j=\frac 52$) with an interaction Hamiltonian
${\cal H} =-2J{\bf S}_i\cdot {\bf S}_j$  has eigenstates with total spin
$S_{\rm T}= |{\bf S}_i+{\bf S}_j|=  0,1,\ldots 2S$, and energy eigenvalues
$E(S_{\rm T})= |J|S_{\rm T}(S_{\rm T}+1) =0,2|J|,6|J|,\ldots,30|J|$, 
as shown in the inset to Fig.~\ref{fcs1}. It is known \cite{Furr79} that
inelastic neutron scattering processes are associated with transitions between
adjacent eigenstates (the selection rules permit $\Delta S_{\rm T} =\pm1$).
The neutron energy gain or loss may thus take values
$\Delta E= \pm2|J|,\pm 4|J|,\ldots,10|J|$. The accuracy of $J$ determination
depends only on the precision of neutron energy transfer measurements.
In fact, the inelastic neutron scattering method is the most direct way of
determining the exchange coupling constants in diluted systems 
\cite{Shap02}. 

\begin{figure*}[t]
\includegraphics[width=6.5in]{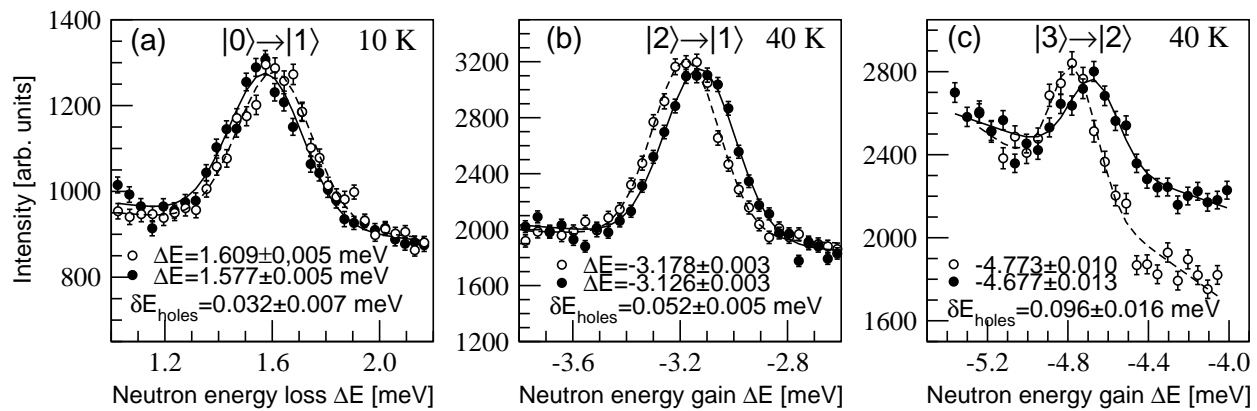}
\caption{\label{fcs2} Inelastic neutron scattering maxima: (a) transition
$|0\rangle \rightarrow |1\rangle$ at 10~K for annealed (full points) and
non-annealed (open points) Zn$_{0.95}$Mn$_{0.05}$Te:P samples; (b,c)
transitions $|2\rangle \rightarrow |1\rangle$ and
$|3\rangle \rightarrow |2\rangle$, respectively at 40~K for annealed (full 
points) Zn$_{0.95}$Mn$_{0.05}$Te:P and undoped reference (open points)
Zn$_{0.98}$Mn$_{0.02}$Te samples.}
\end{figure*} 

A general view of inelastic neutron scattering spectra obtained for our
samples is displayed in Fig.~\ref{fcs1}. The peak on the right side of the
strong elastic line (neutron energy loss scattering process) corresponds to
the $|0\rangle \rightarrow |1\rangle$ transition from the ground state to the
first excited level. The peak intensity visibly decreases when temperature
rises from 10 to 40~K due to the diminished ground state population. Owing to
the low incident energy, only one inelastic line occurs in the energy loss
side of the spectrum. On the energy gain (negative 
$\Delta E$) side  there is a strong $|1\rangle \rightarrow |0\rangle$
transition line and a very weak $|2\rangle \rightarrow |1\rangle$ line at 10~K.
The intensity of the latter increases strongly at 40~K, and a third line 
corresponding to $|3\rangle \rightarrow |2\rangle$ emerges. The broad feature
underneath is most likely caused by phonon scattering. These peaks are shown
in an expanded scale in Fig.~\ref{fcs2}, to make the differences between the
results for particular samples visible. We note that all the inelastic peak
positions measured on as-grown (prior to annealing) sample agree within
experimental error with those observed on the reference sample. 
In contrast, the presence of the holes leads to a small but clearly detectable
shift of the transition energies to lower values, without affecting the
lineshape or scattering intensity. The decrease in the magnitude of the
exchange constant for the sample with the higher hole concentration
demonstrates the ferromagnetic character of the hole-induced contribution 
$J_1^h$. 

The  $|0\rangle \rightarrow |1\rangle$ transition peaks obtained for the
as-grown and annealed Zn$_{0.95}$Mn$_{0.05}$Te:P sample are shown in
Fig.~\ref{fcs2}$(a)$. The peak positions obtained by fitting Gaussian
lineshapes and a sloped background are: $1.609 \pm 0.005$ meV and $1.577 \pm 
0.005$ meV, respectively. This corresponds to the $J_1^h=0.016 \pm 0.004$ meV.
Shown in Fig.~\ref{fcs2}$(b)$ and $(c)$ are peaks corresponding to transitions
$|2\rangle \rightarrow |1\rangle$ and  $|3\rangle \rightarrow 
|2\rangle$, respectively. Here the data for the annealed sample is compared
with the respective data for the undoped reference sample.  For the above two
transitions the shift in the peak positions due to the hole-induced 
ferromagnetic interaction is now expected to increase by a factor of 2 and 3,
respectively. The observed shift values are $0.052 \pm 0.004$ and
$0.096 \pm 0.016$ meV respectively, so that the average value of $J_1^h$
obtained from the full set of the peaks is $0.015 \pm 0.003$ meV. 

Since transport measurements \cite{Ferr01,Khoi02} show that the p-type sample
in question remains on the insulator side of the MIT, we begin the
interpretation of our findings by evaluating the expected magnitude of the 
hole-induced energy shift if the holes were strongly localized at individual
acceptor sites. From earlier studies of bound magnetic 
polarons in p-type DMSs \cite{Jaro85}, we know that the Mn spins residing at
the distance $r$ from the acceptor experience a molecular field,
\begin{equation}
H^*(\bm r) =r_{so}\beta\exp(-2r/a_B)/(2\pi a_B^3g\mu_B),
\end{equation}
where $r_{so}=0.8$ is the reduction factor due to spin-orbit interaction;
$\beta = 0.062$ eVnm$^3$ is the p-d 
exchange integral describing the kinetic exchange interaction between the hole
and Mn spins in Zn$_{1-x}$Mn$_x$Te \cite{Twar84}; $a_B$ is the acceptor Bohr
radius, and $g=2.0$ is the Mn Land\'e factor. Due to the exponential 
decay of the interaction strength, each Mn ion remains under the influence of
a single hole, only so that the expected mean shift of the maximum of the
$|0\rangle \rightarrow |1\rangle$ line reads,
\begin{equation}
\delta E = r_{so}\beta\int d\bm r p g\mu_B H^*(\bm r) = r_{so}\beta
p/2 = 0.12\mbox{~meV}.
\end{equation}
for the hole concentration $p=5 \times 10^{18}$ cm$^{-3}$.
This value is greater by a factor of four than the experimental shift,
$\delta E = 2J_1 = 0.030 \pm0.006$~meV. Furthermore, for the 
Bohr radius \cite{Ferr01}
$a_B = 1.3$~nm, large fluctuations in the value of $\delta E$ 
are to be expected, from zero up to 2.4~meV for a Mn pair located nearby an
acceptor. No evidence for such a doping-induced line broadening is seen in
our results summarized in Fig.~\ref{fcs2}.

We turn, therefore, to those models which assume that -- owing to the
proximity of the metal-insulator transitions -- the holes are weakly localized.
According to the scaling theory of the Anderson-Mott MIT \cite{Beli94}, at
distances smaller than the localization radius (which diverges right at the
MIT), the states retain a metallic character. 
Therefore, in order to evaluate the effect of the holes on the NN Mn pairs,
we refer to the RKKY theory, which gives the carrier-induced exchange energy
between Mn spins located at the distance $r$ in the well known form 
\cite{Diet97,Ferr01},
\begin{equation}
J_1^h(r) = \frac{A_F r'_{so}k_F^4 m^* \beta^2}{4\pi^3\hbar^2}\left
[\frac{\sin(2k_Fr)-2k_Fr\cos(2k_Fr)}{(2k_Fr)^4}\right].
\end{equation}
Here, $A_F = 1.2$ is the Fermi liquid parameter describing the enhancement
of the interaction by the hole-hole exchange \cite{Jung99Koni01};
$r'_{so} =0.48$ is the reduction coefficient due to the spin-orbit interaction 
\cite{Ferr01}; $m^*$ is the effective mass and $k_F$ is the Fermi wavevector.
Since for the NN pair, $r_{NN}k_F \ll 1$, we obtain,
\begin{equation}
J_1^h(r_{\text{NN}}) = A_F r'_{so}\beta^2m^*_tp/(8\pi\hbar^2r_{\text{NN}}),
\end{equation}
where $m^*_t =0.77m_o$ is the sum of the heavy and light hole masses in ZnTe.
For these parameters, we arrive at
$J_1^h(r_{\text{NN}}) =0.010$~meV, which is seen to be somewhat smaller
than the
experimental value $0.015\pm 0.003$~meV. We should note at this point that
the magnitude of $A_F$ may be enhanced on the insulator side of the MIT.
Moreover, three effects conspire in causing the underestimation of the hole
concentration by the Hall resistance measurements: the  anomalous Hall effect,
the Hall scattering factor, and the presence of two-carrier transport (heavy
and light holes). In light of these arguments, we regard the agreement between
the experimental and theoretical values of the hole-induced 
exchange energy as satisfactory. 

In conclusion, our inelastic neutron scattering data have provided direct
experimental information on the character of electronic states mediating
ferromagnetic coupling in a p-type II-VI ferromagnetic semiconductor. The data 
demonstrate that the correct description of the results is possible in terms
of the Zener/RKKY theory, even for the hole concentration as low as
$5\times 10^{18}$~cm$^{-3}$, which corresponds to the insulator side of
the MIT. 

Work supported by the National Science Foundation grant DMR-0204105.
The work in Poland was supported by the by State Committee for Scientific
Research as well as by AMORE project (GRD1-1999-10502) within 5th Framework
Programme of European Commission. One of us (J.K.F.) acknowledges the support
of the National Science Foundation Grant DMR01-38195.

\end{document}